# An Intrinsic Bond-Centered Electronic Glass with Unidirectional Domains in Underdoped Cuprates

Y. Kohsaka[1], C. Taylor[1], K. Fujita[1,2], A. Schmidt[1], C. Lupien[3], T. Hanaguri[4], M. Azuma[5], M. Takano[5], H. Eisaki[6], H. Takagi[2,4], S. Uchida[2,7], J. C. Davis[1,8]*

[1] Laboratory of Atomic and Solid State Physics, Department of Physics, Cornell University, Ithaca, NY 14853, USA.

[2] Department of Advanced Materials Science, University of Tokyo, Kashiwa, Chiba 277-8651, Japan.

[3] Département de Physique, Université de Sherbrooke, Sherbrooke, QC J1K 2R1, Canada.

[4] Magnetic Materials Laboratory, RIKEN, Wako, Saitama 351-0198, Japan.

[5] Institute for Chemical Research, Kyoto University, Uji, Kyoto 601-0011, Japan.

[6] National Institute of Advanced Industrial Science and Technology, Tsukuba, Ibaraki 305-8568, Japan.

[7] Department of Physics, University of Tokyo, Bunkyo-ku, Tokyo 113-0033, Japan.

[8] Condensed Matter Physics and Materials Science Department, Brookhaven National Laboratory, Upton, NY 11973, USA

*To whom correspondence should be addressed. E-mail: jcdavis@ccmr.cornell.edu

**Removing electrons from the $CuO_2$ plane of cuprates alters the electronic correlations sufficiently to produce high-temperature superconductivity. Associated with these changes are spectral weight transfers from the high energy states of the insulator to low energies. In theory, these should be detectable as an imbalance between the tunneling rate for electron injection and extraction−a tunneling asymmetry. We introduce atomic-resolution tunneling-asymmetry imaging, finding virtually identical phenomena in two lightly hole-doped cuprates: $Ca_{1.88}Na_{0.12}CuO_2Cl_2$ and $Bi_2Sr_2Dy_{0.2}Ca_{0.8}Cu_2O_{8+\delta}$. Intense spatial variations in tunneling asymmetry occur** **primarily at the planar oxygen sites; their spatial arrangement forms a Cu-O-Cu bond centered electronic pattern without long range order but with $4a_0$-wide unidirectional electronic domains dispersed throughout ($a_0$: the Cu-O-Cu distance). The emerging picture is then of a partial hole-localization within an intrinsic electronic glass evolving, at higher hole-densities, into complete delocalization and highest temperature superconductivity.**

Metallicity of the cuprate $CuO_2$ planes derives (*1*) from both oxygen $2p$ and copper $3d$ orbitals (Fig. 1A). Coulomb interactions lift the degeneracy of the relevant *d*-orbital, producing lower and upper *d*-states separated by the Mott-Hubbard energy $U$ (Fig. 1B). The lower *d*-states and oxygen *p*-state become hybridized yielding a correlated insulator with charge transfer gap $\Delta$ (Fig. 1B). The "hole-doping" process, which generates highest temperature superconductivity, then removes electrons from the $CuO_2$-plane, creating new hole-like electronic states with predominantly oxygen $2p$ character (*2*). This is a radically different process than hole-doping a conventional semiconductor because, when an electron is removed from a correlated insulator, the states with which it was correlated are also altered fundamentally. Numerical modeling of this process (*3*) indicates that when $n$ holes per unit cell are introduced, the correlation changes generate spectral weight transfers from both filled and empty high-energy bands−resulting in the creation of $\sim 2n$ new empty states just above the chemical potential $\mu$ (Fig. 1B). But precisely how these spectral-weight transfers result in cuprate high-temperature superconductivity remains controversial.

Recently, it has been proposed that these doping-induced correlation changes might be observable directly as an asymmetry of electron tunneling currents with bias voltage (*4, 5*) −electron extraction at negative sample bias being strongly favored over electron injection at positive sample bias. Such effects should be detectable with a scanning tunneling microscope (STM). The STM tip-sample tunneling current is given by

$$I(\vec{r},z,V) = f(\vec{r},z)\int_0^{eV} N(\vec{r},E)\mathrm{d}E \qquad (1)$$

where $z$ is the tip's surface-normal coordinate, $V$ is the relative sample-tip bias, and $N(\vec{r},E)$ is the sample's local-density-of-states (LDOS) at lateral locations $\vec{r}$ and energy $E$. Unmeasureble effects due to the tunneling matrix elements, the tunnel-barrier height, and $z$ variations from electronic heterogeneity are contained in $f(\vec{r},z)$ (see supporting online text 1). For a simple metallic system where $f(\vec{r},z)$ is a featureless constant, Eq. 1 shows that spatial mapping of the differential tunneling conductance $\mathrm{d}I/\mathrm{d}V(\vec{r},V)$ yields $N(\vec{r},E=eV)$. However, for the strongly correlated electronic states in a lightly hole-doped cuprate the situation is much more complex. In theory (4), the correlations cause the ratio $Z(V)$ of the average density-of-states for empty states $\overline{N}(E=+eV)$ to that of filled states $\overline{N}(E=-eV)$ to become asymmetric by an amount

$$Z(V) \equiv \frac{\overline{N}(E=+eV)}{\overline{N}(E=-eV)} \approx \frac{2n}{1+n} \qquad (2)$$

Spectral-weight sum rules (5) also indicate that the ratio $R(\vec{r})$ of the energy-integrated $N(\vec{r},E)$ for empty states $E > 0$ to that of filled states $E < 0$ is related to $n$ by:

$$R(\vec{r}) \equiv \frac{\int_0^{\Omega_c} N(\vec{r},E)\mathrm{d}E}{\int_{-\infty}^{0} N(\vec{r},E)\mathrm{d}E} = \frac{2n(\vec{r})}{1-n(\vec{r})} + O\left(\frac{nt}{U}\right) \qquad (3)$$

Here $t$ is in-plane hopping rate and $\Omega_c$ satisfies "all low-energy scales" $< \Omega_c < U$.

As a test of such ideas, we show in Fig. 1C the predicted evolution of the tunneling asymmetry (TA) with $n$ from (4), while in Fig. 1D we show the measured evolution of spatially-averaged TA in a sequence of lightly hole-doped $Ca_{2-x}Na_xCuO_2Cl_2$ samples with different $x$. We see that the average TA is indeed large at low $x$ and diminishes rapidly with increasing *x,* as predicted (*3-5*). But because such effects are detectable by spatially resolved techniques (*6, 7*), the TA proposals (*4, 5*) also identify the first atomic-scale probe of the doping-induced correlation changes.

**Electronic "cluster glass" state of lightly hole-doped cuprates**

Figure 2A shows schematically that cuprate antiferromagnetism disappears at a doped-hole-density per $CuO_2$ $n \sim 2$ to 3%. The superconductivity usually does not appear until $n \sim 5$ to 10%. Therefore, another low-temperature state intervenes; it is usually thought of as an electronic "cluster glass" (ECG) state (*8-21*). At higher dopings, the ECG signatures coexist with diminishing intensity with the strengthening superconductivity, until they disappear somewhere near $n \sim 15\%$. Although the ECG state exhibits no known long range spin or charge order, some electronic order of unknown spatial form is always detected at nm scale by local probes of the spin (*8-15*) and charge (*14-20*) – hence the "cluster" designation. And, because it is from this ECG state that high temperature superconductivity emerges with hole doping (*22*), it is critical to determine what it is, how it is generated by hole doping, and how it evolves into and coexists with the superconducting state.

Much is known about the cuprate ECG state. Direct evidence for hole-localization comes from (i) in-plane dc resistivity, which exhibits logarithmically increasing temperature dependence $\rho_{ab} \propto -\ln T$ (*23,24*); and (ii) Hall-number measurements showing that the delocalized hole density approaches the chemically doped values only when $n \geq 10\%$ (*25*). Muon-spin rotation measurements (*8-13*) find glassy dynamics of spins but exhibiting some unknown form of spatial order at the nm scale: The spin component of the ECG "cluster". Similarly torque magnetometer studies show (*26*) substantial free-spin paramagnetism. Nuclear magnetic/quadrupole resonance measurements (*14-20*) reveal static charge heterogeneity at the nm scale: The hole-density component of the ECG "cluster". Powder neutron diffraction reveals related local lattice distortions (*21*) and optical spectroscopy the anomalous anisotropic conductivity and electron-phonon couplings (*27*). Finally, neutron-scattering measurements indicate that the holes are clustered in nm-sized regions with some form of magnetic short-range order (*28*). In summary, the cuprate ECG state is pervasive, exhibiting partial hole localization in a state without long-range order but that, nonetheless, supports some unknown form of electronic domains. A long-standing problem has been whether these effects are an intrinsic element of hole-doped $CuO_2$ electronic structure or are extrinsic - perhaps triggered by random dopant and/or impurity disorder.

Theoretical hypotheses for the cause and structure of the cuprate ECG state include, for example, loss of electronic translational invariance because of (i) an electronic glass caused by random exchange couplings (*29*), (ii) a dopant disorder-induced random electronic glass (*30*), (iii) a spontaneous (*31*) or dopant-induced (*30, 32*) glass of self-organized electronic nanodomains, and (iv) a nematic electronic liquid crystal of such nanodomains (*33*). But a direct test of such ideas has not been possible because neither the real-space electronic structure of the ECG state, nor of an individual "cluster", could be determined directly as no suitable imaging techniques existed.

**Design of TA studies in $Ca_{1.88}Na_{0.12}CuO_2Cl_2$ and $Bi_2Sr_2Dy_{0.2}Ca_{0.8}Cu_2O_{8+\delta}$**

STM-based imaging might appear an appropriate tool to address such issues. But d$I$/d$V$ imaging is fraught with problems in lightly doped cuprates. For example, a standard d$I$/d$V$ image although well defined, is not a direct image of the LDOS (see supporting online text 1). Moreover, there are theoretical concerns that, in $Ca_{2-x}Na_xCuO_2Cl_2$, the topmost $CuO_2$ plane may be in an "extraordinary" state (*34*), or that interference between two tunneling trajectories through the $3p_z$-Cl orbitals adjacent to a dopant Na+ ion may cause rotational symmetry breaking in the tunneling patterns (*35*).

The new proposals (*4, 5*) for tunneling asymmetry measurements provide a notable solution to problems with standard d$I$/d$V$ imaging because Eqs. 2 and 3 have a crucial practical advantage. If we define the ratios $Z(\vec{r},V)$ and $R(\vec{r},V)$ in terms of the tunneling current

$$Z(\vec{r},V) \equiv \frac{\frac{dI}{dV}(\vec{r},z,+V)}{\frac{dI}{dV}(\vec{r},z,-V)} \quad (4a) \qquad R(\vec{r},V) \equiv \frac{I(\vec{r},z,+V)}{I(\vec{r},z,-V)} \quad (4b)$$

we see immediately from Eq. 1 that the unknown effects in $f(\vec{r},z)$ are all canceled out by the division process. Thus, $Z(\vec{r},V)$ and $R(\vec{r},V)$ not only contain important physical information (*4, 5*) but, unlike $N(\vec{r},E)$, are also expressible in terms of measurable quantities only. We have confirmed that the unknown factors $f(\vec{r},z)$ are indeed canceled out in Eq. 4 (see supporting online text and figures 2).

To address the material specific theoretical concerns (*34, 35*), we have designed a sequence of identical TA-imaging experiments in two radically different cuprates: strongly underdoped $Ca_{1.88}Na_{0.12}CuO_2Cl_2$ (Na-CCOC; critical temperature $T_c \sim 21$ K) and $Bi_2Sr_2Dy_{0.2}Ca_{0.8}Cu_2O_{8+\delta}$ (Dy-Bi2212; $T_c \sim 45$ K). As indicated schematically in Fig. 2, B and C, they have completely different crystallographic structure, chemical constituents, and dopant species and sites in the termination layers lying between the $CuO_2$ plane and the STM tip. Na-CCOC has a single $CuO_2$ layer capped by a perfectly square CaCl layer and with Na dopant atoms substituted at the Ca site. Dy-Bi2212 has a $CuO_2$ bilayer, above which are both BiO and SrO layers whose unit cells undergo the incommensurate crystal supermodulation; non-stoichiometric oxygen dopant atoms are located interstitially near the BiO layer. Therefore, we assert that TA-imaging phenomena that are identical in these two materials should be ascribed to their only common characteristic – the intrinsic electronic structure of the $CuO_2$ plane.

**Atomic resolution TA-imaging**

In Fig. 2, B and C, we show standard d$I$/d$V$-spectra measured under identical junction conditions at random locations on the surfaces of the Na-CCOC and Dy-Bi2212 samples (all data were acquired at 4.2 K). Within $|E| < 100$ meV, they both exhibit the expected V-shaped d$I$/d$V$ centered on $E = 0$ (*6, 7*). Unexpectedly, at higher energies, the same intense spatial variations in the tunneling asymmetry of spectra were observed in both materials. These can be seen vividly at the left hand perimeter of Fig. 2, B and C (because our procedures normalize the integrated d$I$/d$V$ on the positive side). The corresponding variations in TA indicate the existence of intense atomic-scale variations in electronic structure.

To explore the spatial arrangements of these phenomena, we used an atomic resolution "$R$ map" - spatially imaging $R(\vec{r},V)$ of Eq. 4b. Figure 3, A and B, show typical topographic images of the CaCl and BiO layers obtained by cleavage, in cryogenic ultrahigh vacuum, of Na-CCOC and Dy-Bi2212, respectively. The brightest regions in Fig. 3A indicate the locations of Cl atoms that are directly above the Cu atoms in Na-CCOC, whereas those in Fig. 3B indicate the Bi atoms that are above the Cu atoms in Dy-Bi2212. The dark, cross-shaped regions in Fig. 3A are the missing Cl atoms and, in Fig. 3B, are displaced Bi atoms along maxima of the crystalline supermodulation. Figures 3, C and D are images of $R(\vec{r},V = 150\,\mathrm{mV})$ measured in the identical fields of view of Fig. 3, A and B, respectively. These $R$ maps are markedly similar in texture and exhibit far finer spatial details than their related surface topographs; the reason is that much of their contrast stems from features occurring within each Cu plaquette (Fig. 1A). The $R$ maps exhibit no long range spatial order of any kind. Nevertheless, autocorrelation analysis shows that they do have short-range ~ 4$a_0$ × 4$a_0$ periodic correlations, where $a_0$ is the Cu-O-Cu distance. The most obvious and arguably most important observation in Fig. 3, C and D, is a loss of both translational and 90°-rotational (C$_4$) invariance in the spatial arrangements of electronic structure at the 4$a_0$ scale – these effects being virtually indistinguishable in Na-CCOC and Dy-Bi2212. It is also evident from Fig. 3, C and D, that the internal structure of these "domains", as well as the overall matrix in which they are embedded, retains further degrees of electronic complexity at the atomic scale.

**Cu-O-Cu bond-centered electronic glass with disperse 4$a_0$-wide domains**

To visualize these spatial elements more clearly, we take the Laplacian $\nabla^2 R$ of Fig. 3,

C and D (Fig. 3, E and F). At atomic scale, we then see an electronic structure consisting of $a_0$-length elements distributed in a disordered fashion along both Cu-O directions. Within this matrix are embedded $4a_0$-wide unidirectional regions or "domains". These domains, because they are periodic along the long axis, appear to be ordered. Repeating $4a_0$-wide domains of this type are always unidirectional, extending along one or other Cu-O direction. Thus, at the nm scale, the electronic structure of these lightly hole-doped cuprates breaks both $C_4$ symmetry and translational symmetry of the ideal square crystal lattice.

In Fig. 4, A and D, we show higher resolution studies of equivalent domains from Na-CCOC and Dy-Bi2212, respectively (at the fine blue boxes of Fig. 3, C and D). Here the $R$ maps are rotated to put a CuO axis, and thus the domain axis, vertical. The pairs of dark lines (representing high asymmetry) in $R$ maps indicated by the arrows are precisely $4a_0$ apart and represent the perimeter of a single domain. In Fig. 4, A and D, multiple, parallel, $4a_0$-wide domains run from the bottom to the top of each image - exhibiting virtually identical internal structure in both materials.

We next examine, in Fig. 4, B and E, the internal structure of the domains (at the boxes of Fig. 3, A and B, and Fig. 4, A and D) with identification of atomic sites from the simultaneous topographs (Fig. 4, C and F, respectively). We see immediately that the primary spatial variations in the $R$ maps are concentrated, not on the Cu sites, but rather on the O site within each Cu-O-Cu bond. Here the domain's symmetry axis is along a vertical line starting at the arrows labeled 1. Along this axis are a line of oxygen sites, each within a horizontal Cu-O-Cu bond and all exhibiting high $R$. The vertical line labeled 2 is the line of vertical Cu-O-Cu bonds; these oxygen sites exhibit low $R$. Thus, $R$ is very different for the horizontal Cu-O-Cu bonds transverse to line 1 and the vertical Cu-O-Cu bonds along line 2, even though these bonds share a Cu atom on the corner of the same plaquette. The next vertical line of Cu atoms away from the axis is labeled 3, and line 4 represents the line of oxygen sites that is $2a_0$ to the right of axis 1. The sequence of horizontal Cu-O-Cu bonds along line 4 exhibits a uniformly low $R$. These patterns exhibit mirror symmetry about the vertical axis 1 - meaning that the whole domain is precisely $4a_0$ wide. We find these uniaxial domains in all *R* maps – randomly dispersed with equal probability of orientation along the two Cu-O axes (Fig. 3, E and F) and with virtually identical structure in both materials.

A noteworthy observation here is that the O sites within Cu-O-Cu bonds, even though crystallographically equivalent, are in electronically inequivalent states (Figs. 3 and 4). In general, the spatial arrangements of these Cu-O-Cu bond states exhibit no long range order. Nonetheless, there are clear short range relations between them: Sequential vertical or horizontal Cu-O-Cu bonds along a vertical axis can all be in same state, whereas the Cu-O-Cu bonds at 90° to each other and sharing a corner Cu atom are electronically inequivalent. Most notably, these TA images indicate that the cuprate electronic "cluster glass" (*8-21*) stems from spatial variations in the electronic state of each Cu-O-Cu bond.

**Atomic-scale electronic structure within the $4a_0$-wide domains**

Next we consider the energy dependence of electronic structure of the domains in Fig. 4. Although d*I*/d*V* images are not simply related to spatial arrangements of LDOS, individual d*I*/d*V* spectra still retain much physical importance – especially in the energy value at which their key features occur. To clarify the locations, relative to the O and Cu orbitals in the $CuO_2$ plane, of each d*I*/d*V* spectrum measured on the surfaces of Fig. 4, C and F, we show a schematic in Fig. 5A. The d*I*/d*V* spectra are measured along equivalent lines labeled 1,2,3, and 4 in both domains of Fig. 4, B and E, and Fig. 5A. The nine spectra (a to i) along lines labeled 1 and 4 are taken at a sequence of five planar oxygen sites with four empty sites in between, whereas those along lines 2 and 3 are taken at a sequence of five planar copper sites with four oxygen sites in between. In both materials, they reveal similar spatial evolutions: Line 1 shows spectra with minimal TA and clear low energy features (at ~10 meV in Na-CCOC and ~20 meV in Dy-Bi2212); line 2 exhibits rapid TA fluctuations but low-energy features similar to those of line 1; line 3 shows higher TA; and line 4 show the highest TA with weak low energy features. These patterns of d*I*/d*V* spectra exhibit mirror symmetry about the domain's vertical axis. All spectra show a pronounced feature at $E \sim$ +100 meV with a related feature near $E \sim$ -100 meV but masked by the rapid rise due to tunneling asymmetry. The low-energy features, whether of peaks (line 1 and 2) or shoulders (line 3 and 4), appear at $\sim \pm 10$ meV for Na-CCOC and $\sim \pm 20$ meV for Dy-Bi2212, with a clear reduction of d$I$/d$V$ towards $E = 0$. These types of spectral shapes, and the different energy ranges in samples with different $T_c$, may indicate their relation to the superconductivity.

**Long-range electronic structure**

Returning to the largest scales, the $R$ map shown in Fig. 6 spans a 25-nm$^2$ field of view. No long range order can be detected (even out to 50 nm; fig. S3). But the Fourier transform of such $R$ maps (inset to Fig. 6) reveals further surprises: The predominant peaks occur at wavevectors $\vec{q} \sim (3/4,0)$ and $(0,3/4)$ (in units of $2\pi/a_0$), whereas the peaks at $\vec{q} \sim (1/4,0)$ and $(0,1/4)$ that would be expected trivially from a $4a_0 \times 4a_0$ modulation are weaker. We find that these characteristic TA modulations at $\vec{q} \sim (3/4,0)$ and $(0,3/4)$ occur not because of mixing between $(1/4,0)$ and $(1,0)$, but primarily because three maxima in $R$ typically exist within each $4a_0$-wide domain

**Discussion and conclusions**

Because we find virtually identical phenomena in Na-CCOC and Dy-Bi2212 samples of radically different physical, chemical, and dopant structure, material-specific explanations (*34*, *35*) for the TA effect can be ruled out. Instead, we consider these phenomena to be intrinsic electronic characteristics of the CuO$_2$ plane. Furthermore, the effects reported here cannot be governed by individual dopant atoms because there is only a single dopant atom for every ~20 Cu-O-Cu bonds and, in any case, they occur at quite different locations in the unit cells of Na-CCOC and Dy-Bi2212. Similarly, the creation by random dopant distributions of virtually identical unidirectional $4a_0$-wide electronic domains in both materials (Fig. 4) appears extremely unlikely. Indeed, the spatial arrangements of electronic structure revealed by TA imaging (Figs. 3, 4, and 6), rather than exhibiting a random patchy configuration as expected if dopant disorder predominates, occur because of variations in the electronic state at each Cu-O-Cu bond. Such a bond-centered electronic glass, if intrinsic and universal to cuprates, would provide a plausible and consistent explanation for why long range spin/charge-ordered states are not detected at low doping.

Tunneling asymmetry measurements are designed to yield key information on how correlations affect electronic structure of the hole-doped cuprates (*4,5*). But the primary spatial contrast in TA images, although undoubtedly electronic, has not yet been independently calibrated for the degree of charge-density variations it represents and may also contain quantum interference effects (*35*). In addition, the relation between the higher -energy *R* maps and the low energy d*I*/d*V*-maps is incompletely understood. For example, Fourier transforms of d*I*/d*V* images at |*E*| < 50 meV reveal "checkerboard" d*I*/d*V* modulations with peaks at $\vec{q}$ ~ (0,3/4) and (3/4,0) (*6*). This implies two, at present indistinguishable, possibilities about the relation between *R* maps and d*I*/d*V* maps: (i) The physical entity of the bond-centered domains may appear differently in the two experimental quantities (see supporting online text 1); and/or (ii) the localized electronic domains affect delocalized low-energy states, possibly via scattering interference (*36*). But independent of which explanation for "checkerboard" d*I*/d*V* modulations holds, these new TA-imaging techniques are well defined, avoid the systematic errors from $f(\vec{r},z)$ in d*I*/d*V* imaging, access higher-energy scales than previously, and may reveal insightful new physical information (*3-5*).

Because the TA images always contain $4a_0$-wide unidirectional patterns, an obvious question is whether they are segments of the charge/spin-ordered "stripes" (*37-41*). It is argued from neutron scattering that $4a_0$-wide unidirectional long-range charge order forms the basis of static stripes (*42, 43*). Consistent with this picture, resonant X-ray scattering in $La_{1.875}Ba_{0.125}CuO_4$ reveals $4a_0$-periodic modulations of hole density at oxygen sites (*44*). But the complete atomic-scale electronic structure of a cuprate "stripe" is unknown because direct spectroscopic imaging has been unachievable. Our TA data on $4a_0$-wide unidirectional domains (Figs. 3, 4, and 5) seem consistent with the experimental understanding (*42-44*) of the "stripes" in La-based cuprates - except that here there is no long range order [consistent with results from (*45*)]. A possible explanation is that a Cu-O-Cu bond-centered electronic glass, although ubiquitous in cuprates at low hole-doping, may be converted to long-range static "stripe" order due to unique crystal symmetry and commensuration in $La_{2-x}Ba_xCuO_4$ and $La_{1.6-x}Nd_{0.4}Sr_xCuO_4$ at $x = 1/8$. In any case, direct detection of $4a_0$-wide unidirectional electronic domains by STM represents an exciting opportunity to determine the internal electronic structure of a cuprate "stripe" (Fig. 4). Furthermore, because our data appear consistent with the "cluster" phenomenology (*8-21,23-25*), the $4a_0$-wide electronic domains also represent excellent candidates to be the ubiquitous "clusters" of the ECG state

Finally, our samples also exhibit the tenuous *d*-wave superconductivity of lightly hole-doped cuprates. This coexists spatially with the bond-centered tunneling asymmetry patterns described here. *A priori* the TA contrast represents variations in the ratio of probability of electron extraction to injection – implying atomically varying probability of electronic occupancy. By contrast, the superconductivity consists of delocalized *d*-wave electron pairs. How such disparate effects can coexist at low doping, and how the full delocalization (*25*) associated with highest-temperature superconductivity emerges from this state with increased doping, remain to be determined.

**Figure captions**

Fig. 1

(A) Relevant electronic orbitals of the $CuO_2$ plane: Cu $3d$ orbitals are shown in orange and oxygen $2p$-orbitals are shown in blue. A single plaquette of four Cu atoms is shown within the dashed square box, and a single Cu-O-Cu unit is within the dashed oval.

(B) Schematic energy levels in the $CuO_2$ plane and the effects of hole-doping upon it.

(C) The expected tunneling asymmetry between electron extraction (negative bias) and injection (positive bias) from (*4*) where low values of $Z$ occur at low hole densities $n$.

(D) Measured doping dependence of average tunneling asymmetry in $Ca_{2-x}Na_xCuO_2Cl_2$. a.u., arbitrary units.

Fig. 2

(A) Schematic phase diagram of hole-doped cuprates. The regions of antiferromagnetic insulator (AFI), $d$-wave superconductor ($d$SC), and pseudogap (PG) phenomenology are indicated. The electronic "cluster glass" (ECG) extends over a region indicated beneath the dashed line.

(B and C) Examples of the strong spatial variations in the tunneling asymmetry found in Na-CCOC and Dy-Bi2212. Because experimental normalization keeps integrated $dI/dV$ at positive biases constant, the large variations are seen at negative bias and directly reflect spatial variation of the TA (tunnel junction set at 200 pA, 600 mV). As can be seen, electron extraction is strongly favored over injection and varies spatially. These phenomena are quite similar in the two materials.

Fig. 3

(A and B) Constant-current topographic images of Na-CCOC and Dy-Bi2212 in 12-nm$^2$ fields of view. Imaging conditions are (A) 50 pA at 600 mV and (B) 50 pA at 150 mV. The orange boxes in (A) and (B) indicate areas described in Fig. 4, B and C, and Fig. 4, E and F, respectively, and the Cu-O bond directions are shown as pairs of orthogonal black arrows.

(C and D) $R$ maps taken at 150 mV [i.e., $R(\vec{r},150\,\mathrm{mV}) = I(\vec{r},+150\,\mathrm{mV})/I(\vec{r},-150\,\mathrm{mV})$] in the same field of view shown in Fig. 3, A and B, respectively. Large $R$ (bright in this color scale) means that the corresponding tunneling spectrum is more symmetric, whereas low $R$ (dark) means that it is more asymmetric. The blue boxes in (C) and (D) indicate the boxed areas of Fig. 4, A and D, respectively.

(E and F) Images of $\nabla^2 R$ (Laplacian) computed from Fig. 3, C and D, respectively, for better visualization of the atomic-scale arrangements of the spatial patterns.

Fig. 4

(A and D) $R$ maps of Na-CCOC and Dy-Bi2212, respectively (taken at 150 mV from areas in the blue boxes of Fig. 3, C and D). The fields of view are (A) 5.0 nm by 5.3 nm and (B) 5.0 nm by 5.0 nm. The blue boxes in (A) and (D) indicate areas of Fig. 4, B and C and Fig. 4, E and F, respectively.

(B and E) Higher resolution $R$-map within equivalent domains from Na-CCOC and Dy-Bi2212 respectively (blue boxes of Fig. 4, A and D). The locations of the Cu atoms are shown as black crosses.

(C and F) Constant-current topographic images simultaneously taken with Fig. 4, B and E, respectively. Imaging conditions are (C) 50 pA at 600 mV and (F) 50 pA at 150 mV. The markers show atomic locations, used also in Fig. 4, B and E. The fields of view of these images are shown in Fig. 3, A and B, as orange boxes.

Fig. 5

(A) Locations relative to the O and Cu orbitals in the $CuO_2$ plane where each $dI/dV$ spectrum at the surfaces of Fig. 4, C and F, and shown in Fig. 5B, is measured. Spectra are measured along equivalent lines labeled 1, 2, 3, and 4 in both domains of Fig. 4, B and E, and Fig. 5A.

(B) Differential tunneling conductance spectra taken along parallel lines through equivalent domains in Na-CCOC and Dy-Bi2212. All spectra were taken under identical junction conditions (200 pA, 200 mV). Numbers (1 to 4) correspond to trajectories where these sequences of spectra were taken. Locations of the trajectories relative to the domains, are shown between Fig. 4, B (C) and E (F) by arrows.

Fig. 6

A 25-nm$^2$ $R$ map; no long-range order is apparent. Instead, we see randomly distributed electronic variations of the Cu-O-Cu bond state with equal probability of orientation along the two Cu-O axes. The Cu-O bond directions are shown as pairs of orthogonal black arrows. The inset shows its Fourier transform. The predominant peaks occur at wavevectors $\vec{q}$ ~ (3/4,0) and (0,3/4) in units $2\pi/a_0$ (orange arrows), and the peaks at $\vec{q}$ ~ (1/4,0) and (0,1/4) (blue arrows) are weaker. Atomic peaks $\vec{q}$ ~ (1,0) and (0,1) are shown by black arrows.

**Fig. 1**

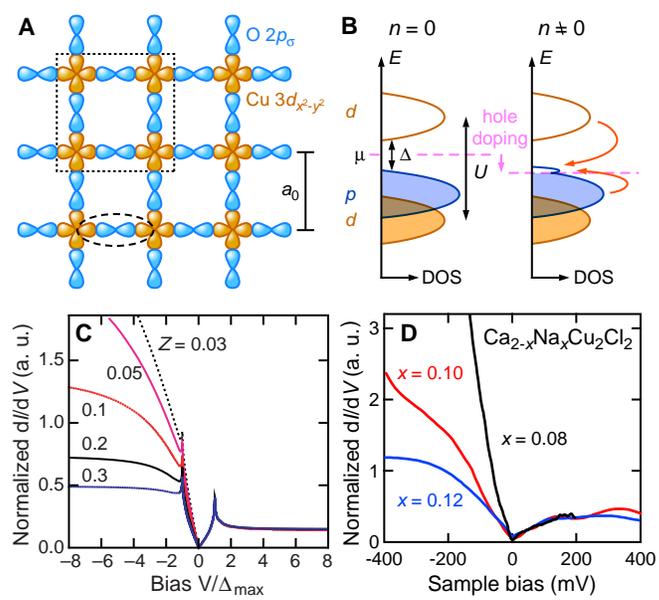

**Fig. 2**

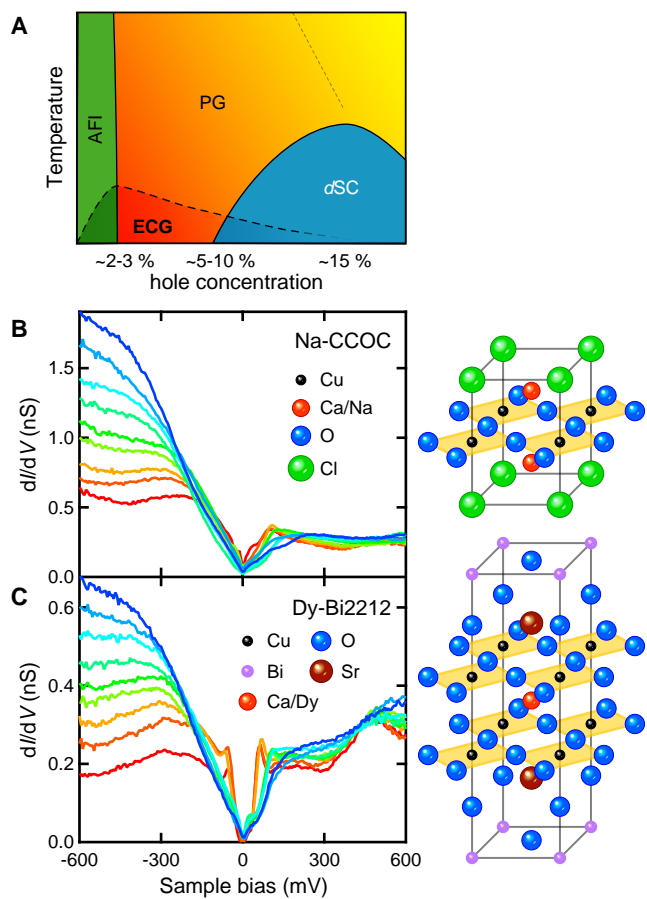

**Fig. 3**

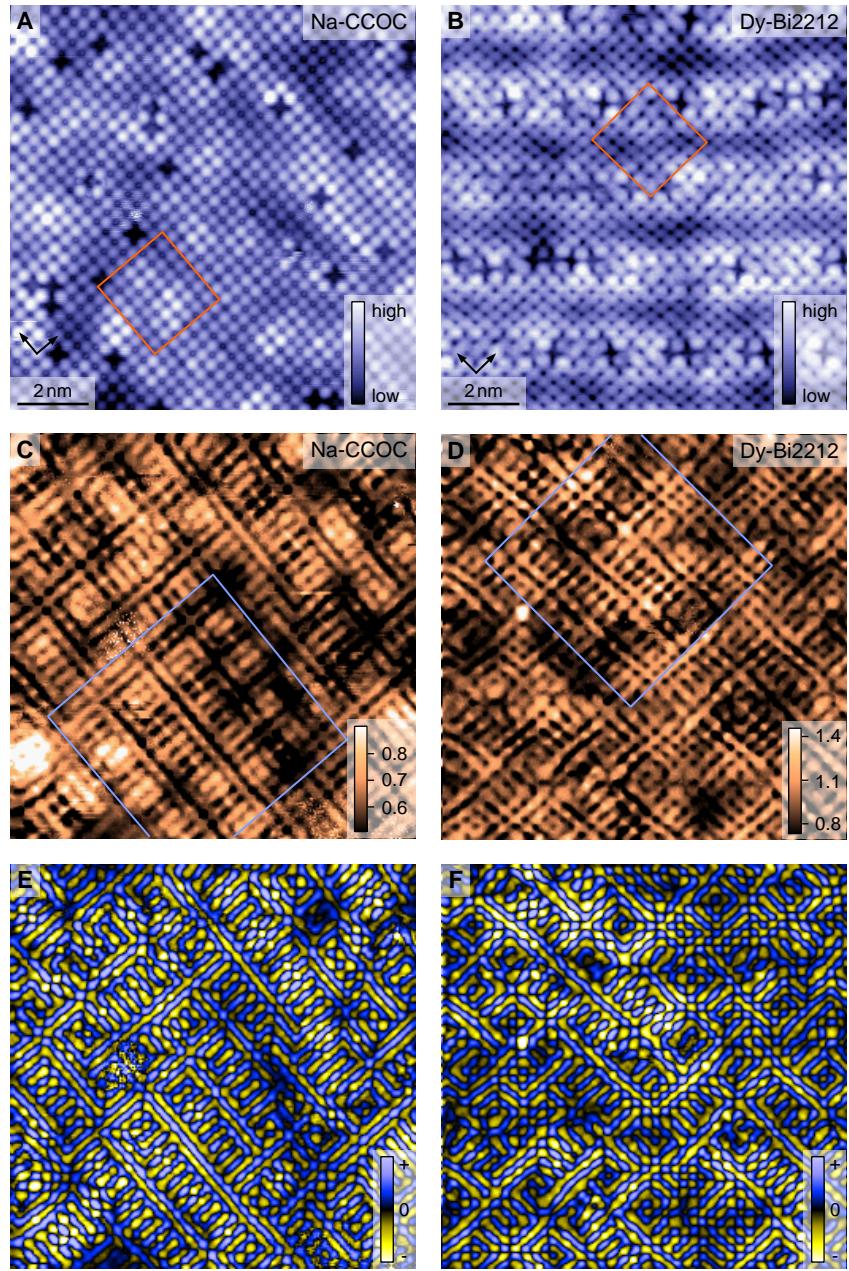

**Fig. 4**

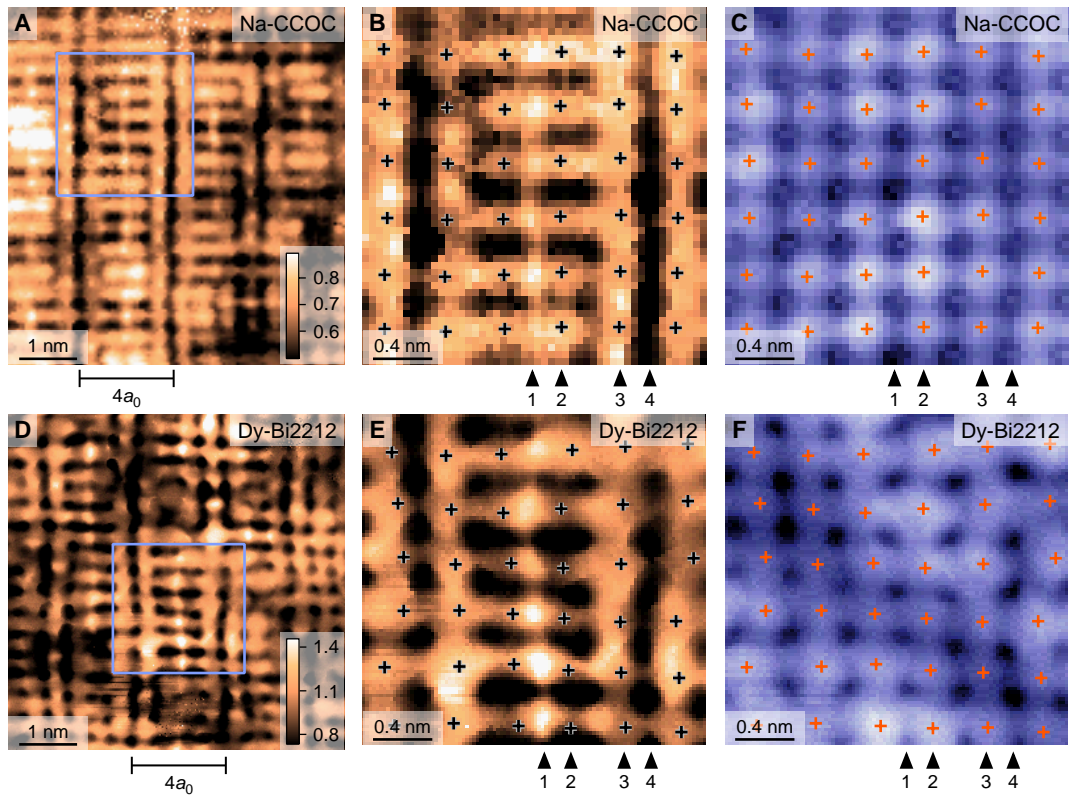

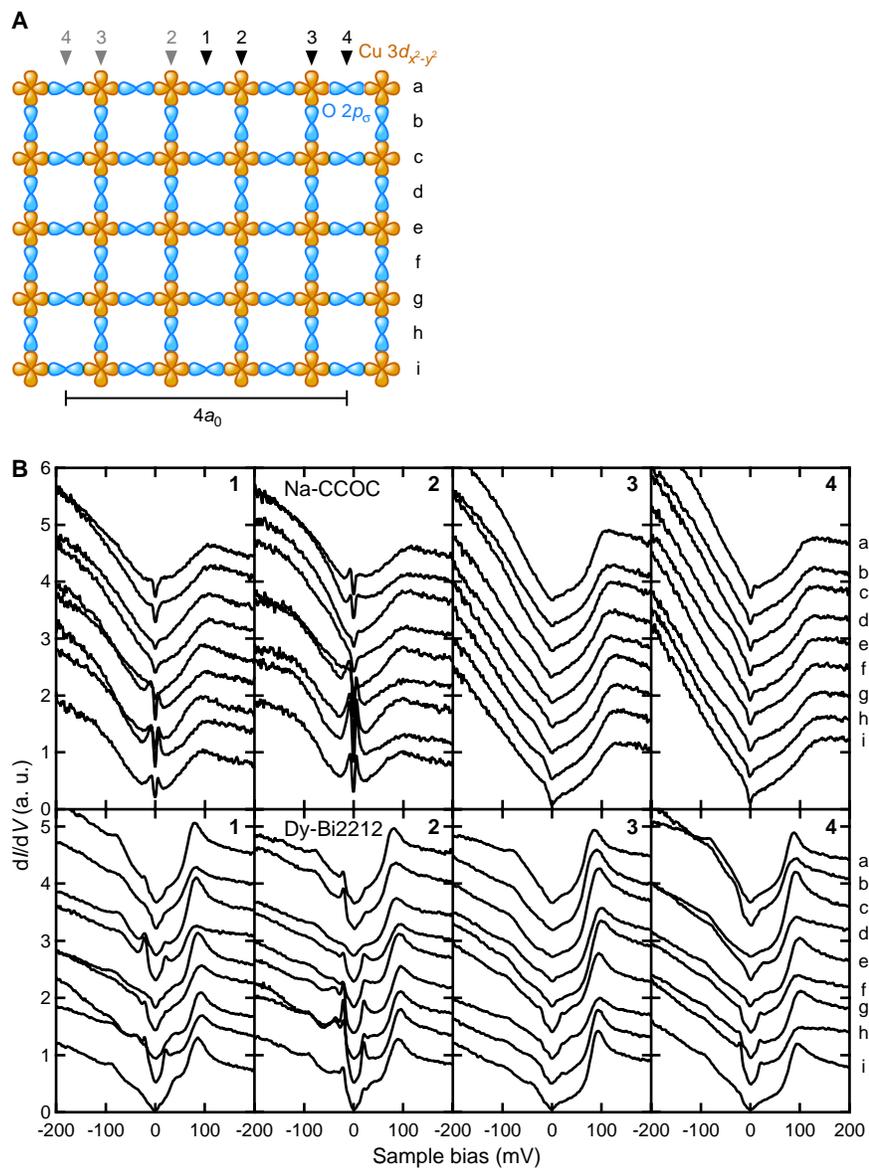

**Fig. 6**

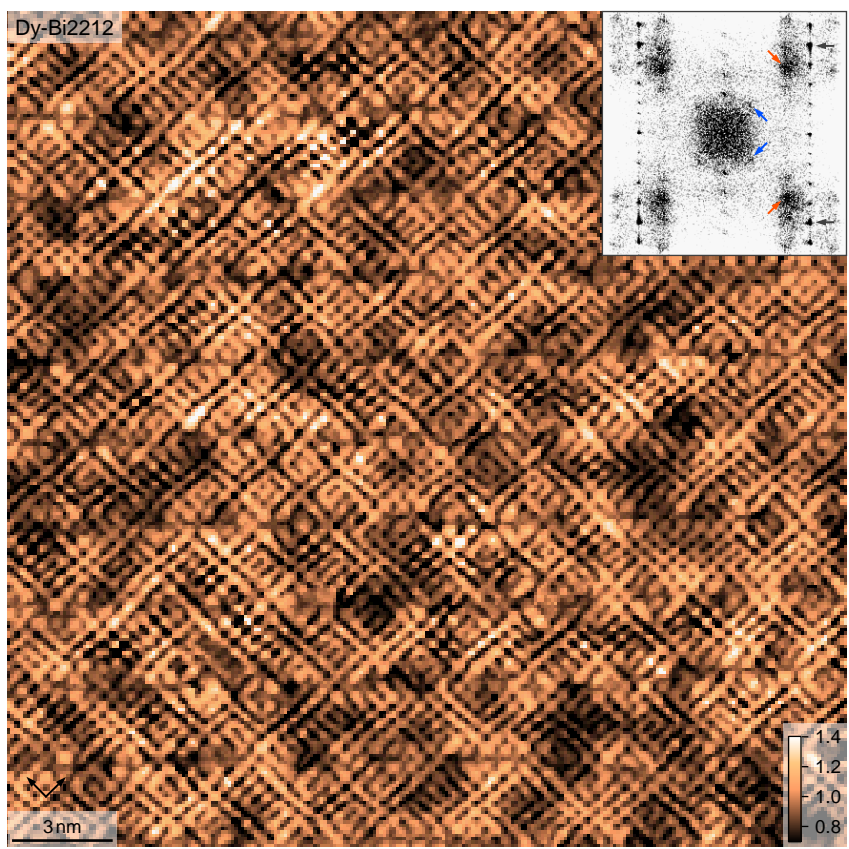